\documentclass[conference]{IEEEtran}
\IEEEoverridecommandlockouts
\usepackage{cite}
\usepackage{amsmath,amssymb,amsfonts}
\usepackage{algorithmic}
\usepackage{graphicx}
\usepackage{textcomp}
\usepackage{xcolor}
\usepackage{balance}
\usepackage{makecell}
\usepackage{hyperref}
\usepackage{url} 
\usepackage{listings} 
\usepackage{float}
\usepackage[skip=5pt]{caption}
\usepackage[utf8]{inputenc}

\usepackage[letterpaper,%
            left=0.75in,right=0.75in,top=0.75in,bottom=1in,%
            footskip=.25in,bindingoffset=0.2in]{geometry}

\def\BibTeX{{\rm B\kern-.05em{\sc i\kern-.025em b}\kern-.08em
    T\kern-.1667em\lower.7ex\hbox{E}\kern-.125emX}}

\usepackage{graphicx} 

\begin{document}

\makeatletter
\newcommand{\linebreakand}{%
 \end{@IEEEauthorhalign}
 \hfill\mbox{}\par
 \mbox{}\hfill\begin{@IEEEauthorhalign}
}
\makeatother

\title{A Case Study in Gamification for a Cybersecurity Education Program: A Game for Cryptography



} 
\author{

   \IEEEauthorblockN{Dylan Huitema}
   \IEEEauthorblockA{\textit{Computer Science} \\
   \textit{Okanagan College}\\
    Kelowna, Canada \\
   0009-0002-8931-8578}

 \and

   \IEEEauthorblockN{Albert Wong}   \IEEEauthorblockA{\textit{Mathematics and Statistics} 
  \\   \textit{Langara College}\\
   Vancouver, Canada \\
   0000-0002-0669-4352}
   }

\maketitle
\begin{abstract}
 Advances in technology, a growing pool of sensitive data, and heightened global tensions has increased the demand for skilled cybersecurity professionals. Despite the recent increase in attention given to cybersecurity education, traditional approaches have continue in failing to keep pace with the rapidly evolving cyber threat landscape. Challenges such as a shortage of qualified educators and resource-intensive practical training exacerbate these issues.

 Gamification offers an innovative approach to provide practical hands-on experiences, and equip educators with up-to-date and accessible teaching tools that are targeted to industry-specific concepts. The paper begins with a review of the literature on existing challenges in cybersecurity education and gamification methods already employed in the field, before presenting a real-world case study of a gamified cryptography teaching tool. The paper discusses the design, development process, and intended use cases for this tool. This research highlights and provides an example of how integrating gamification into curricula can address key educational gaps, ensuring a more robust and effective pipeline of cybersecurity talent for the future.

\end{abstract}


\section{Introduction}
The advancement of artificial intelligence (AI) tools, the growing pool of sensitive data on individuals and institutions and governments, and the ever-increasing global tensions have exacerbated the need for the production of skilled cybersecurity professionals. As the cyber threat landscape has evolved in recent years, there has been a growing concern within the cybersecurity community that training and education have not kept pace\cite{Armstrong2020, GISWS2017}. The introduction of gamification concepts within cybersecurity curricula could play a crucial role in bridging this gap\cite{Malone2021}.

For many reasons, an effective cybersecurity education is difficult to provide in practice. Significant efforts are required to integrate practical hands-on exercises into a curriculum, which is difficult, as this method of instruction requires more resources and specialized equipment than a regular lecture or lab. The lack of qualified educators, researchers, and experts also hinders effective program development. Even with appropriate content and qualified educators, continuous changes in cybercriminal tactics and the rapid adoption of new technology require near-constant updates to this content\cite{Ferrari2023}. The rate of change should not be expected to slow down anytime soon. With our increased connectedness through the Internet of Things (IoT) networks, the rapid improvement in AI, and cloud services, the incentive for cybercrime is at an all-time high. All of this places greater strain on what many would say is an already struggling education system.  

This is not to say that the problem has not been addressed; it is just that the problem may require new solutions. In fact, cybersecurity education within traditional computer science programs has seen increased emphasis and significant advancement. Various solutions have been proposed \cite{Malone2021, Scherb2023, Taylor-Jackson2020}. The use of technologies such as gaming and AI has been used to improve student engagement and promote deeper understanding \cite{DeBello2023, Alansi2023}. The number of programs available is also growing rapidly, especially in terms of postsecondary education and online certifications. 

At the K-12 level, cybersecurity is a difficult area to teach, but the current situation makes it clear that it is necessary\cite{Chen2021}. A key issue here is that many computer teachers in elementary and even high schools do not have postsecondary education specific to computer science or cybersecurity. The gap here is twofold: first, there was little to no cybersecurity education available pre-university; and second, many teachers now tasked with teaching computer concepts may lack the understanding needed to effectively demonstrate key concepts\cite{Ibrahim2024}. The large-scale implementation of educational tools, such as games, has the potential to be instrumental in closing this gap by providing these teachers with an up-to-date, practical, and engaging way to approach topics that may not otherwise be addressed. 

The main contribution of this paper is to provide a tangible and applied example of gamification that is directly related to core computer science topics. Rather than examining theoretical gamification ideas, this paper discusses their practical implementation of a ready-to-use educational tool. The presented example could be used as a building block to work toward the presentation of many cybersecurity concepts, specifically cryptography, in a new and innovative cybersecurity prgram. 

\section{Existing Works}
This section reviews existing cybersecurity education methods, gamification techniques that have been introduced, and proposed solutions through peer-reviewed articles published in the last three years. The review focuses on identifying the challenges, methods, and outcomes of cybersecurity education at all levels, especially those relevant to the development of innovative educational tools such as gamification. 

A common challenge outlined in cybersecurity education is the difficulty of designing a comprehensive curriculum, largely due to the field's rate of change and the need for practical, hands-on instruction. Although universities around the world are increasingly aware of the need for dedicated cybersecurity programs, their development and the resources required remain a significant barrier. Ismail et al. published a survey on activities used in cybersecurity education, mainly in the UAE, a country where cyber attacks have intensified the focus on cybersecurity education. Their findings emphasize that hands-on activities provide students with a greater understanding of cryptographic principles. A particularly relevant example was their discussion of a gamified exercise on the Diffie-Hellman key exchange, the structure of which is very similar to the levels in the game discussed later in this paper.

Also in 2024, Ibrahim et al.\cite{Ibrahim2024} published a systematic review of K-12 cybersecurity education around the world. Their main conclusion is that cybersecurity is not widely included in international curricula in a meaningful way. They instead found that rather than structured instruction, students often develop awareness through personal device use at home or in unsupervised situations. They concluded that cybersecurity education is also far from standardized, with some including it as a branch of computer science, while others treat it as its own discipline. 

In contrast, considerably more attention has been paid to cybersecurity education methods within higher education. Ferrari et al. \cite{Ferrari2023} in 2023 reviewed the literature on cybersecurity education within higher education institutions. Their research highlights the difficulties in developing effective cybersecurity programs and advocates for the need to do so. They also discuss various recent innovations and frameworks designed to improve curricula, such as gamification and the integration of AI tools. 

The potential of infusing cybersecurity education with artificial intelligence has also been explored \cite{DeBello2023, Parambil2024}. Like many other fields, the use of AI in cybersecurity education has the potential to revolutionize the education process. Wang et al. in 2025 \cite{Wang2025} introduce an AI-powered tool called CyberMentor, a large language model (LLM), designed to assist students and instructors by providing mentorship and access to an organized repository of educational resources. They see promising results in testing, especially in practical, skill-based exercises. 

Wagner and Alharthi in 2023 \cite{Wagner2023} discuss several different virtual reality technologies (VR / AR / MR / XR) and their potential to improve cybersecurity education. Their research finds that the use of * R technology can increase both knowledge attainment and retention. As a large portion of these technologies use gamification principles, these findings are directly applicable to this research. Other research, such as Malone et al. \cite{Malone2021}, Karmakar et al. \cite{Karmakar2022}, and Alansi et al. \cite{Alansi2023}, all discuss the potential of gamification to improve levels of knowledge attainment and retention, motivation, and learning outcomes.

The use of gamification concepts has been in use with regard to specific areas of cybersecurity for some time. Games like CTF (Capture the Flag) have been used since the late 90s, to a significant effect. Williams et al. \cite{Williams2024} discuss the use of these games and their role in gaining student interest in cybersecurity in their articles. With an open-ended approach to designing and implementing CTF-adjacent games, they believe traditional cybersecurity education curricula can be greatly enriched and diversified.

The effects of implementing gamification concepts in educational domains other than cybersecurity have also been documented \cite{Lampropoulos2024, Pavlova2022, Cigdem2024, Gue2022}. The potential for improvement in areas such as student interest and retention, the two areas in which cybersecurity education has faced ongoing challenges, is evident throughout the reviewed literature.

In summary, while cybersecurity faces a plethora of unique challenges, research shows that emerging areas like AI integration, gamification concepts, and virtual reality capabilities offer a promising path forward. Hands-on learning should be the focal point in the future, and embracing the methods described can help bridge existing gaps in cybersecurity education.

\section{The Cryptography Game}

In this section, we present the concepts of a game that can be used in a cybersecurity education program. The game's focus is on introducing cryptographic algorithms and their role in dealing with cybersecurity issues. 

\subsection{Design and Development}

In the game,  a student assumes the role of a highly intelligent spy working for an intelligence agency. The game's first level is designed to introduce cryptographic algorithms by decrypting coded messages sent between two opposing operatives. The choice of the player character is based on the fact that an experienced spy equipped with the necessary tools and knowledge is perfect for introducing ideas throughout the game.  

As an example, level 1 begins when the player intercepts a coded message. The spy is then explicitly told by his agency that intelligence suggests that the message is encrypted using a Caesar cipher. The explanation of the Caesar cipher is then presented in the game as the thoughts of the spy. 

 \begin{figure}[H] 
    \centering
    \includegraphics[width=0.5\textwidth]{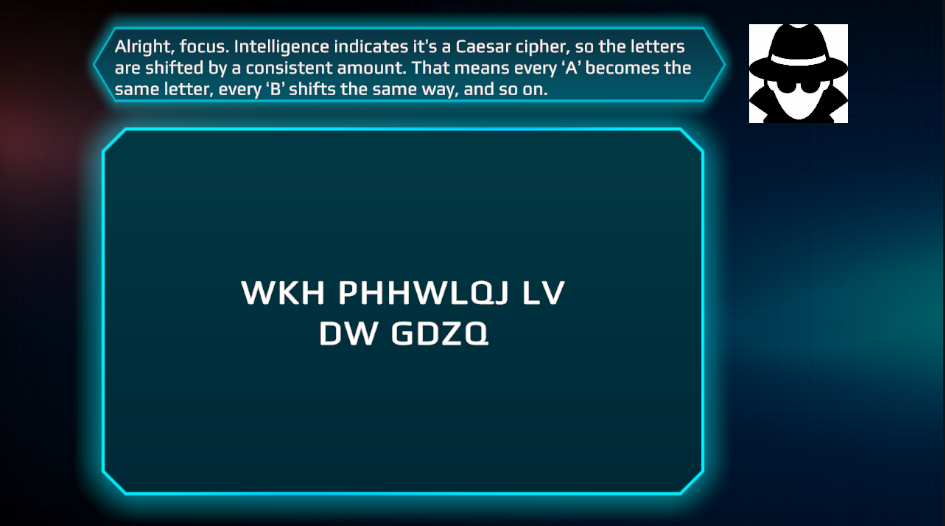}
    \caption{Upon intercepting the message.}
    \label{fig:Level 1 intro} %
\end{figure}
\vspace{-5pt}

The student is then tasked with decrypting the message using arrow keys to determine the shift to apply to the message through a Caesar cipher. Depending on the number of attempts and the time that the player takes to complete the decryption, the player character will provide advice or additional information about the task.

  \begin{figure}[H] 
    \centering
    \includegraphics[width=0.5\textwidth]{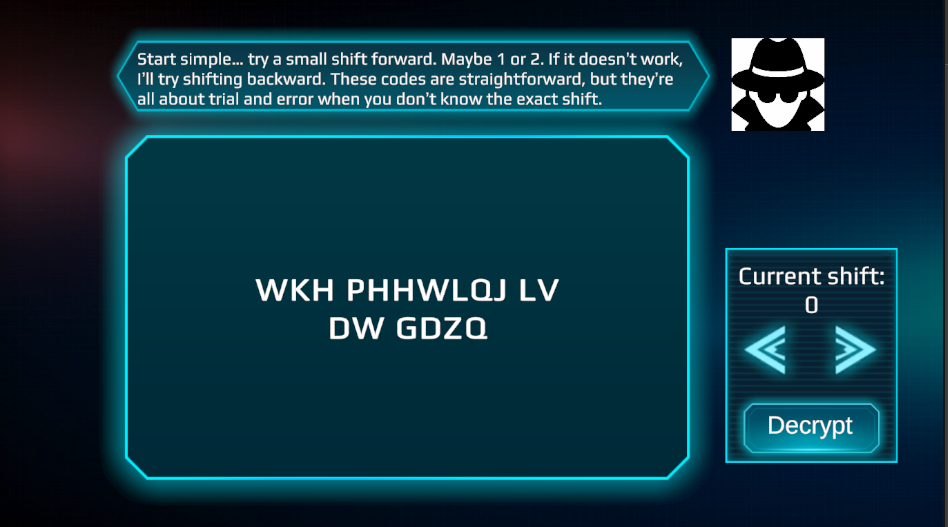}
    \caption{Advice from character on how to solve caesar cipher.}
    \label{fig:Level 1 intro 2} %
\end{figure}
\vspace{-5pt}

As the student attempts to decrypt the message, the player character will provide input as to whether they are moving in the right direction. In the case of a Caesar cipher, the underlying mechanism allows for somewhat little input in terms of advice. However, as the game progresses and the demands on the player increase, the advice becomes a more important inclusion, serving as a helping hand if a student gets stuck. 

 \begin{figure}[H] 
    \centering
    \includegraphics[width=0.5\textwidth]{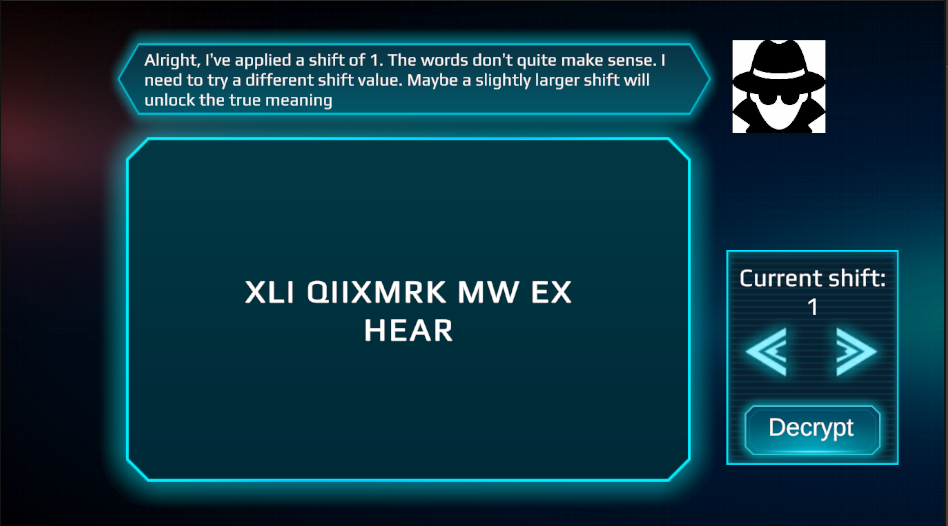}
    \caption{First shift attempt.}
    \label{fig:Level 1 intro 3} %
\end{figure}
\vspace{-5pt}

Upon completing this level, the students are greeted with a success message and their character's thoughts on how the information gained affects the game's larger story. 

 \begin{figure}[H] 
    \centering
    \includegraphics[width=0.5\textwidth]{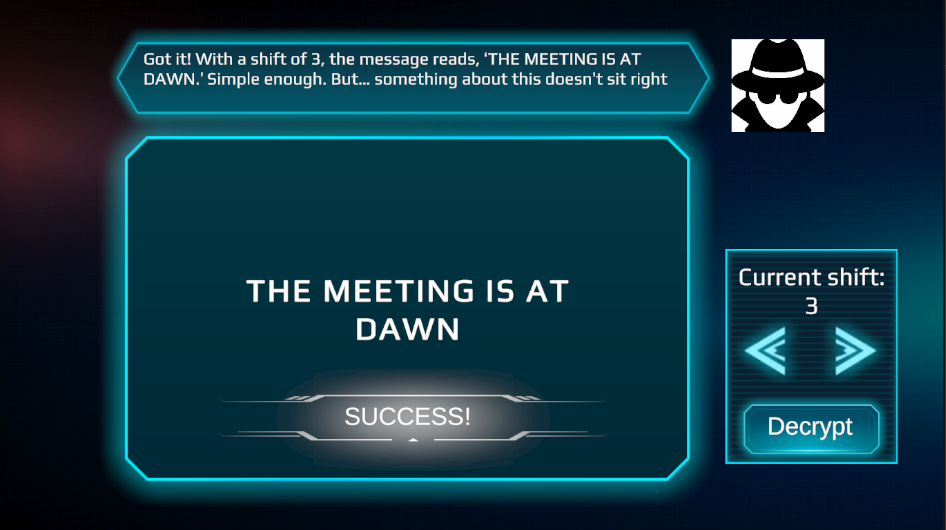}
    \caption{Successful decryption.}
    \label{fig:Level 1 intro 4} %
\end{figure}
\vspace{-5pt}

The final screen, before moving on, is designed to encourage the student to interact with the codex. The codex is a collection of all the concepts the student has been introduced to throughout the game. As the game progresses, the mechanisms needed to solve the puzzles increase. Students are meant to use the codex as a tool to help uncover which of the concepts they have learned is applicable in a given situation. 

 \begin{figure}[H] 
    \centering
    \includegraphics[width=0.5\textwidth]{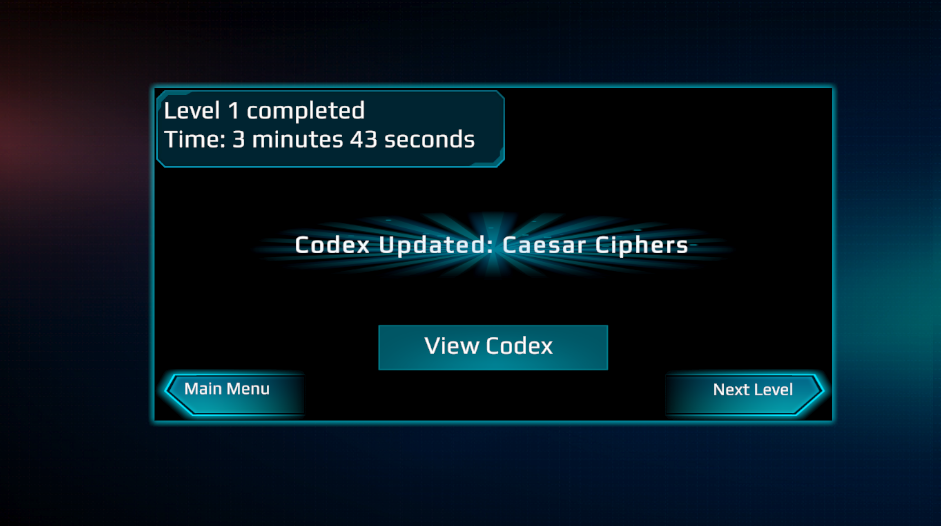}
    \caption{Codex update.}
    \label{fig:Level 1 intro 5} %
\end{figure}
\vspace{-5pt}

The above serves as a simple look at the game's core gameplay loop, even though it is the first and easiest level. As the game progresses, concepts become more abstract, and, as expected when working through any course materials, more is demanded of the students. For instance, in sections later on that cover topics like Block Ciphers and Stream Ciphers, students are required to write code and understand ideas like keys, initialization vectors, and Exclusive OR (XOR) operations to decrypt messages. 

\subsection{Topics Covered}

In deciding which cryptography concepts should be included, a review of typical concepts covered in courses offered during an undergraduate degree was conducted. The review provides a foundation for ensuring that the game aligned with existing educational standards and that key theoretical and practical concepts were covered. However, the emphasis of the review is on finding areas where improvements could be made. The courses looked at included those offered at Okanagan College, the Cryptography and Cryptanalysis program from the University of Southern California, the Cryptography I program offered on Coursera by Stanford University, and various Introduction to Cryptography courses offered at other education institutions. The topics covered in three commonly used textbooks in these courses were also reviewed. An exhaustive list of the courses and materials reviewed can be found here\cite{MATH231, Boneh2021, Reyzin2021, Huang2021, KatzLindell2015, BonehShoup2023}. 

The topics decided on and covered in the game roughly in the same order are presented below.

\begin{enumerate}
    \item \textbf{Classical Cryptography}
    \begin{itemize}
        \item Substitution Ciphers
        \item Transposition Ciphers
    \end{itemize}

    \item \textbf{Symmetric-Key Cryptography}
    \begin{itemize}
        \item Block Ciphers
        \item Stream Ciphers
        \item Key Management
    \end{itemize}

    \item \textbf{Public-Key Cryptography}
    \begin{itemize}
        \item Encryption Algorithms
        \item Elliptic Curve Cryptography
        \item Diffie-Hellman Key Exchange
    \end{itemize}

    \item \textbf{Cryptographic Hash Functions}
    \begin{itemize}
        \item Key Concepts
        \item Common Hash Functions
    \end{itemize}

    \item \textbf{Key Management and Key Distribution}
    \begin{itemize}
        \item Key Exchange Protocols
        \item Key Distribution Systems
        \item Key Escrow
    \end{itemize}

\end{enumerate}

This list is not exhaustive. A cryptography student who focuses on cybersecurity may graduate with a breadth of knowledge beyond this list. However, modern concepts such as Post-Quantam Cryptography, Blockchain and Cryptocurrencies, and Threshold Cryptography are difficult to fully grasp, let alone gamify. Although not ruled out in the future, the scope of these topics is beyond that of this paper. 

In addition, foundational mathematical principles that underpin cryptography puzzles are included in supplementary materials such as the Codex, even though they are not explicitly listed here. These materials are adequately covered in many of the programs and texts reviewed, but the core concept of the game does not align well with their inclusions in the design. 

\subsection{Educational Impact}
This section aims to articulate how a game like the one described above can improve traditional teaching methods. We begin by mapping elements of the game to the gamification concepts. The potential impact of those concepts is then linked to the learning outcomes in a cybersecurity education program. 

The game's first and most apparent advantage is the progress mechanism. Progressing through levels, regardless of how minor the progression is, can significantly impact motivation \cite{Karmakar2022}. For some students, the progression through the game's story will also serve as a motivating factor. 

Another advantage of gamification and the focus on developing the game is the inclusion of immediate feedback. Cryptography, like many areas of Computer Science, is complex and, at times, lacks positive feedback in the learning process. If something works, then great - typically, the program runs, and you get the expected output or result, but nothing really happens. When something goes wrong, you will know it - error messages, infinite loops, unexpected output, etc. Providing a controlled environment for coding exercises like cryptographic algorithms has many advantages in this area. For one, with failed attempts by a student, the game can provide a supportive dialogue. Hints can be offered when students are struggling with a particular concept. On successful attempts, in addition to completing the task, students are rewarded with a message of success, a small piece of the game's story, and access to the next level. for cryptography exercises, the game transforms the traditionally binary feedback into a more interactive and rewarding learning experience.

Based on the above, using the game will result in a higher level of student engagement with cryptography concepts. More importantly, the impact of the widespread adoption of these principles within cybersecurity education can result in a more inviting environment for potential students and a more engaging and rewarding environment for existing ones. 

\section{Conclusion}

The primary purpose of this research is to develop the concepts of a game to teach cryptography. This is to lay the foundation for gamification ideas within a cybersecurity education program, rather than to deliver a fully finished product to be used immediately. The approach focuses on applying the ideas of gamification and centers on a topic that we feel will benefit from its advantages. With that said, there is strong evidence that improvements to the proposed game concepts or the development of similar ones for other concepts in cybersecurity can have a strong positive impact in areas like student retention and interest.

\balance

\bibliographystyle{IEEEtran}

\bibliography{DylanCybersec}

\end{document}